\documentstyle[preprint,prl,aps]{revtex}
\input{epsf.tex}
\input psfig
\include{psfig}
\begin{document}

\title{Evidence of Skyrmion excitations about $\nu =1$
        in n-Modulation Doped Single Quantum Wells by Inter-band Optical
        Transmission}

\author{E. H. Aifer and  B. B. Goldberg}
\address{Dept. of Physics, Boston University, Boston MA 02215}

\author{D. A. Broido}
\address{Physics Department, Boston College, Chestnut Hill, Ma. 02167}

\date{\today}
\maketitle
\begin{abstract}
  We observe a dramatic reduction in the degree of spin-polarization of
  a two-dimensional electron gas in a magnetic field when the Fermi
  energy moves off the mid-point of the spin-gap of the lowest Landau
  level, $\nu=1$.  This rapid decay of spin alignment to an unpolarized
  state occurs over small changes to both higher and lower magnetic
  field. The degree of electron spin polarization as a function of $\nu$
  is measured through the magneto-absorption spectra which distinguish
  the occupancy of the two electron spin states.  The data provide
  experimental evidence for the presence of Skyrmion excitations where
  exchange energy dominates Zeeman energy in the integer quantum Hall
  regime at $\nu=1$.

{PACS numbers:  73.40.Hm, 78.66.-w, 73.20.Mf, 71.70.Gm, 73.20.Dx}
\end{abstract}
\vspace{2mm}
\newpage

The exchange energy dominates the basic physics in GaAs two-dimensional
electron systems (2DES) when the Fermi energy is located between
spin-split Landau levels at odd-integral filling factors $\nu$. This is
because the small $g$-factor in GaAs makes the Zeeman energy much less
than the Coulomb energy which is responsible for the exchange. The small
$g$ results in a spin degree of freedom, even at high magnetic fields,
leading to spin-unpolarized ground-states and to novel excited-states at
fractional filling factors.\cite{halperin-hpa83} Recent theoretical work
has pointed out that the response of a 2DES with a small $g$-factor in the
spin-polarized state ($\nu=1$) to a change of one quanta of magnetic
flux is not a single-particle spin-flip excitation, but rather a
macroscopic spin object called a Skyrmion or charged spin-texture
excitation (CSTE) \cite{sondhi-prb93,fertig-prb94}. Evidence of these
excitations have been recently observed in NMR and in tilted-field transport
measurements\cite{barrett-prl95,eisenstein-95}. They consist of a radial
spin density distribution that is reversed at the center but gradually
heals to the spin background over a distance of many magnetic lengths.
Since each particle sits in a nearly aligned spin neighborhood, the
exchange contribution is significantly smaller for the CSTE than for a
single flipped spin. The physical size of the CSTE is governed by the
relative strength of the Zeeman and Coulomb energies, parameterized by
$\tilde{g} \equiv E_Z/E_C \: = \frac{g \mu_{B}B}{e^{2} / \epsilon l_0}$,
\cite{fertig-prb94} where $l_0= \sqrt{\hbar/eB}$ is the magnetic length
and $\epsilon$ the dielectric constant.  In the limit of vanishing
$g$-factor, the radius extends to nearly the edge of the sample, while for
$\tilde{g} \geq 0.02$ the size shrinks to zero, eliminating the
distinction between single-particle and Skyrmion excitations. In GaAs samples
with 2DES densities of $1.5 \cdot 10^{11} cm^{-2}$, $\tilde{g} \sim
0.015$ at $\nu=1$.  CSTE's are then expected to be the lowest energy
excitations with the change in total spin per flux-quanta significantly
greater than one, destroying the spin-polarization of the electron
system for small excursions from $\nu=1$.

In this letter we present experimental observation of Skyrmions through
the rapid loss of spin-polarization about $\nu=1$ measured with
polarized absorption spectroscopy. The spectra show quenching of
absorption to the lower energy, spin-up electron band directly
correlated to an increase in the higher energy, spin-down absorption at
$\nu=1$ (see figure ~\ref{sky-fig1}). As we will show, this indicates
the spin-up state fills with electrons while the spin-down state
empties, providing a large spin polarization $S_z$ (see inset) which
exhibits a pronounced, symmetric decay when $\nu$ deviates from $1$.
This new technique provides a measurement of the absolute electron
spin and has identified saturation in the spin polarization not previously
resolved.

The samples were two single-side n-modulation doped AlGaAs - GaAs $250
\AA $ single quantum wells (SQW).  Sample A had mobility $\mu = 3.2
\cdot 10^6 cm^2/Vs$ and 2DES carrier concentration of $n = 1.5 \cdot
10^{11}cm^{-2}$, and sample B had $\mu = 2.6 \cdot 10^6 cm^2/Vs$ and $n
= 1.8 \cdot 10^{11}cm^{-2}$. In transport these wafers exhibited strong
fractional Hall minima at $\nu=1/3$ and $2/3$. The SQWs were chosen to
minimize inhomogeneous broadening, and for absorption measurements were
mounted strain-free and thinned to $\sim 0.5 \mu m$\cite{bbg-prb88}.
Incident power of $\sim 1$mW/cm$^2$ yielded typical signal-to-noise
ratios of $>20$ with line widths of $0.2$ to $0.5$ meV (FWHM).  The
absorption coefficients $ \alpha^+, \alpha^-$, were calculated
neglecting reflection which has been measured in similar samples to
contribute only a small variation ($< 5 \% $).  The raw transmission
spectra $I(w,B)$ were then normalized to obtain the magneto-absorption
coefficient $\alpha(w,B) = - 1/L_{w} \ln(I/I_{0})$, where $L_{w} = $
quantum well width.

In this work we concentrate only on the lowest Landau level in the
regime from $\nu = 0.6\ {\rm to}\ 1.4$ about the spin gap.
Representative spectra taken in LCP and RCP polarizations are displayed
in the lower left and right of Figure ~\ref{sky-fig1}, respectively. As
described below, the final electron spin state for the lowest energy LCP
(RCP) absorption is the lower (higher) energy spin-up (spin-down) state.
The inter-band optical absorption is proportional to the available
density of states in these final electron spin levels.  The total
spin-polarization is then given by the difference between the number of
spin-up and spin-down states under the constraint that the sum yields
the particle number.  With this technique we are able to determine
within $10\%$ accuracy the total spin. The inset to figure
{}~\ref{sky-fig1} shows the spin-polarization $S_z$, as a function of
filling factor determined in this way for the data presented.

We should mention that a wealth of data on a seemingly similar effect,
the quenching of the photoluminescence from the lowest energy transition
accompanied by an increase in the emission from the higher energy
transition, has been observed by one of the authors and
others\cite{bbg-prl90}.  The total integrated emission was relatively
constant, yielding an explanation based on a decrease in the
re-combination rate in the lowest energy state due to localization. In
the absence of significant non-radiative channels, the minority
photo-excited hole must eventually emit a photon on re-combination, and
hence the emission from the two electron levels must be correlated.
However, not only is absorption largely unaffected by localization, but
also the photons absorbed which cause transitions into the lower and
higher energy spin states are completely uncorrelated.  This leads to
the conclusion that the correlation observed in the absorption data is
due to the changes in the occupancy and thus the total spin of the
electron system.

To determine the occupancy of the electron spin states and hence the
spin-polarization, we have first calculated the inter-band transitions
and optical matrix elements. Sub-band energies and wave-functions for
electrons and holes were determined self-consistently within the local
density approximation. The hole Landau levels were then calculated
employing the Luttinger Hamiltonian\cite{luttinger-pr56} to take into
account the valence band mixing.\cite{yang-prb85} Figure ~\ref{sky-fig2}
compares the calculated and measured peak energy positions versus
magnetic field about $\nu=1$.  The insets display measured and
calculated spectra, and the lower right schematic identifies the
relevant transitions: The lowest energy transition in RCP is from a pure
heavy-hole state with $m_j=-3/2$ to the upper electron spin state
$m_j=-1/2$ which we label $0H^- \to e^-$. The lowest energy state in LCP
is from a mixed heavy-hole state with dominant character in the $m_j =
+3/2, +1/2$ components to the lower energy spin state $m_j=+1/2$,
labeled $0H^+ \to e^+$.

Note that the $m_j=+1/2$ part of the hole envelope function for
$0H^+ \to e^+$ is associated with a higher oscillator index and cannot
cause an optical transition to the lowest electron Landau
level.\cite{bbg-prb88} This is corroborated both by the absence of
absorption peaks at the same energy in the different polarizations, and the
narrowness (less than the bare Zeeman energy) of the $0H^+ \to e^+$
transition at high fields. These observations confirm the validity of
the matrix element calculations in the axial approximation.

We proceed to determine the spin level occupancy from the raw
data as follows: The spin-polarization per particle is given by
\begin{equation}
  \begin{array}{lcccc}
    S_z & = &  \frac{N_{\uparrow}-N_{\downarrow}}{N} & = &
    \frac{N_{A_{\downarrow}} - N_{A_{\uparrow}}}{N}
    \end{array}
\label{eq-Sz}
\end{equation}

\noindent
where $N_{\uparrow(\downarrow)} = N_B - N_{A_{\uparrow(\downarrow)}}$.
$N_B$ is the Landau level degeneracy $eB/h$, and $N_{A_j}$ is the
available density of states in the $j=\uparrow$ ($\downarrow$) spin up
(down) band of the lowest Landau level. The integrated absorption peaks
are linearly related to the number of available final states $N_{A_j}$
of the transition as
\begin{equation}
  I_{ij} = C \cdot f_{ij} N_{A_j}
\label{eq-NA}
\end{equation}

\noindent
where $I_{ij} = \int \alpha_{ij} d\omega$, the oscillator strengths
$f_{ij}(\omega)$ are taken to be constant over the narrow absorption
peaks, and $i,j$ label the initial and final states respectively.
The constant of proportionality $C$ may be found using the sum-rule
\begin{eqnarray}
   \frac{N_{A_{\downarrow}} + N_{A_{\uparrow}}}{N} & = &
   \frac{(N_B - N_{\uparrow})+(N_B - N_{\downarrow})}{N} \nonumber \\
   & = & \frac{2-\nu}{\nu}
\label{eq-Csum}
\end{eqnarray}

\noindent
which conserves particle number $\frac{N_{\uparrow}+N_{\downarrow}}{N}=1$.
Then
\begin{equation}
  C(B) = \frac{\nu}{2-\nu} \left(
         \frac{I_{i,\uparrow}}{f_{i,\uparrow}} +
         \frac{I_{k,\downarrow}}{f_{k,\downarrow}}
         \right)
\label{eq-C}
\end{equation}

\noindent
and the available densities of states are obtained from (~\ref{eq-NA}).
The calculated scaling factor $C(B)$ changes by less than $15\%$ over
the range of $\nu=0.6\ {\rm to}\ 1.4$, while typical $<S_z>$'s change by
nearly an order of magnitude, demonstrating that the raw data come very
close to obeying the sum-rule over this field range and indicating that
few higher-order processes are affecting the absorption.

Several additional self-consistency checks exist for this treatment,
providing confidence in our hole level and matrix elements calculations.
The calculation of $S_z$ from the $0H^- \to e^-$ and $0H^+ \to e^+$
transitions, and the one from the $2H^- \to e^-$ and $0H^+ \to e^+$
transitions are nearly identical in the range where the transitions have
good signal to noise (see inset to Fig. 1.). Since the $0H^-$ and the
$2H^-$ independently monitor the occupancy of the upper electron
spin state, the similarity of the spin-polarizations calculated means
that the matrix elements are internally consistent with our simple
sum-rule and that the data truly reflect a change in occupancy of the
electron spin states.  Finally, the matrix elements themselves are
varying relatively slowly over the filling factor range of interest
$\nu=0.6\ {\rm to}\ 1.4$, typically less than $20\%$, and hence cannot
simply account for the structure observed. Nor are they particularly
sensitive to the carrier density or the precise value of the zero-field
splitting; these parameters have been varied with no significant change
in the final spin polarization $S_z$.

In figure ~\ref{sky-fig3} spin polarization versus filling factor is
plotted for samples A and B and compared with both a single particle and
a Skyrmion based model. The single particle model is based on exchange
enhanced $g$-factor that modulates the overlap of the
two electron spin levels \cite{ando-jpsj74}. $g$ is the
self-consistent solution of
\begin{equation}
  g = g_0 + \frac{\epsilon_{xc}^0(N_{\uparrow}-N_{\downarrow})}{\mu_B B}
\label{eq-1p}
\end{equation}

\noindent
where
\begin{equation}
  N_{\uparrow (\downarrow)} = \frac{2B}{\Gamma\sqrt{\pi}}
  \int f(E,E_f) \cdot e^{-\left(\frac{E \pm g\mu_B B/2}{\Gamma/2}
  \right)^2} dE
\label{eq-gsc}
\end{equation}

\noindent
$f(E,E_f)$ is the Fermi distribution function, and $\Gamma = \Gamma_0
\sqrt{B}$ is the field dependent level width. The exchange coefficient
$\epsilon_{xc}^0$ in (~\ref{eq-1p}), was chosen to satisfy $g=7.3$ at
$\nu=1$, as determined from activated transport measurements in
\cite{usher-prb90}, leaving $\Gamma_0$ the only adjustable parameter.
Clearly the single particle model does not capture the behavior of
$S_z$. The polarization quickly saturates to unity for $\nu < 1$ and
goes as $\sim \frac{2-\nu}{\nu} = N_{A_{\downarrow}}/N$ for $\nu > 1$,
in contrast to the measured polarization which decays symmetrically about
$\nu=1$ at a much more rapid rate.

The changes in polarization do adhere however, to a treatment which
includes Skyrmion excitations (see solid line fit
figure ~\ref{sky-fig3}). In the model proposed by Barret {\em et al.}
\cite{barrett-prl95} the one-particle available densities of states are
scaled by a parameter $S$ ($A$) that gives the number of spin flips per
unpaired flux quanta $|N_B - N|$, above (below) $\nu=1$. In this model
the spin polarization is

\begin{equation}
  S_z = \left\{
    \begin{array}{lr}
      S \left( \frac{2-\nu}{\nu} \right) - (S-1) & \nu > 1  \\
      \frac{1}{\nu} - (2A-1) \left( \frac{1-\nu}{\nu} \right)
      & \nu < 1
    \end{array}
  \right.
\label{eq-Sz-sk}
\end{equation}

\noindent
Particle-hole symmetry requires that the size of the Skyrmion be the
same as the Anti-Skyrmion, $S = A$, giving a rapid quasi-symmetric loss
in polarization about $\nu=1$ for $S >1$. When $S=A=1$ the single
particle model is recovered (modulo overlap effects).  The fit for
sample B in figure ~\ref{sky-fig3} gives a skyrmion size of of $3.7$ which
is near the theoretically predicted value for a 2DES in GaAs of $3.5$
\cite{sondhi-prb93,fertig-prb94}. For sample A the skyrmion size is
somewhat smaller, only $2.5$ flipped particles per flux quanta.

A feature of our technique is that it allows a quantitative determination
of the total spin. The data display a marked saturation in the peak spin
polarization at $\nu=1$ for decreasing temperature (see figures
{}~\ref{sky-fig3} and ~\ref{sky-fig4}). The saturation could be due to the
finite level width with the result of non-vanishing overlap of the
spin-states at $\nu=1$, consistent with our measured line-widths.

In conclusion, we have demonstrated a novel method of extracting spin
polarization from interband absorption spectroscopy. Self-consistent
checks of the subband and matrix element calculations as well as
an adherence of the raw data to a simple sum rule provide confidence in
our technique. The measured size of the charged spin-texture excitations
is consistent with a spin of $\sim 3$ flips per unpaired flux quanta.

This work was supported by National Science Foundation grant
DMR-9158097. We are also thankful for helpful discussions with
S. Barrett.

\bibliographystyle{aip}

\vfill\eject

 \section{Figure captions}

 \noindent {\bf Figure ~\ref{sky-fig1}:} Absorption spectra in LCP and
 RCP in the neighborhood of $\nu=1$. The quenching of absorption to the
 lower spin state is directly correlated to an increase in the
 absorption to the upper spin state. The calculated spin-polarization is
 plotted as a function of filling factor $\nu$ using the lowest energy
 LCP transition for the spin-up occupancy and data from both the RCP
 (solid line) and LCP (dashed line) transitions for the spin-down state
 occupancy.

\noindent {\bf Figure ~\ref{sky-fig2}:} The energy of the two lowest
transitions (in LCP and RCP) to the ground Landau level are plotted as a
function of magnetic field and compared to the calculations. The
transitions are identified in the lower left, and spectra and
calculations at 12T are plotted versus energy in the upper insets. Note
that while some discrepancy exists in the absolute energy position, the
calculated matrix elements capture very closely the strength of the
optical transitions.

\noindent {\bf Figure ~\ref{sky-fig3}:} The  calculated
spin-polarization $S_z$ displayed as a function of filling factor $\nu$
for $1.4$ and $0.5$K. The solid line fit assumes a macroscopic spin of
$\sim 3.7$ per flux quanta, while the dashed, dotted, and dash-dot fits
assume a single particle self-consistent exchange-enhanced $g$-factor
model with appropriate ranges of broadening and temperature.

\noindent {\bf Figure ~\ref{sky-fig4}:} Spin-polarization as a function
of temperature in sample A with a carrier density of $n=1.5\times
10^11$cm$^-2$. The peak of the spin-polarization increases to $0.8$ for
decreasing temperature where it saturates. Most notable is the increase
in width of the region of spin-polarization, which may be due to a
relatively wide $\nu=1$ integral Hall plateau in this sample, and the
resultant effect of increased carrier localization on local exchange.
\vfill\eject

%
%
%
\newbox\figa
\setbox\figa=\vtop{\kern0pt\psfig{figure=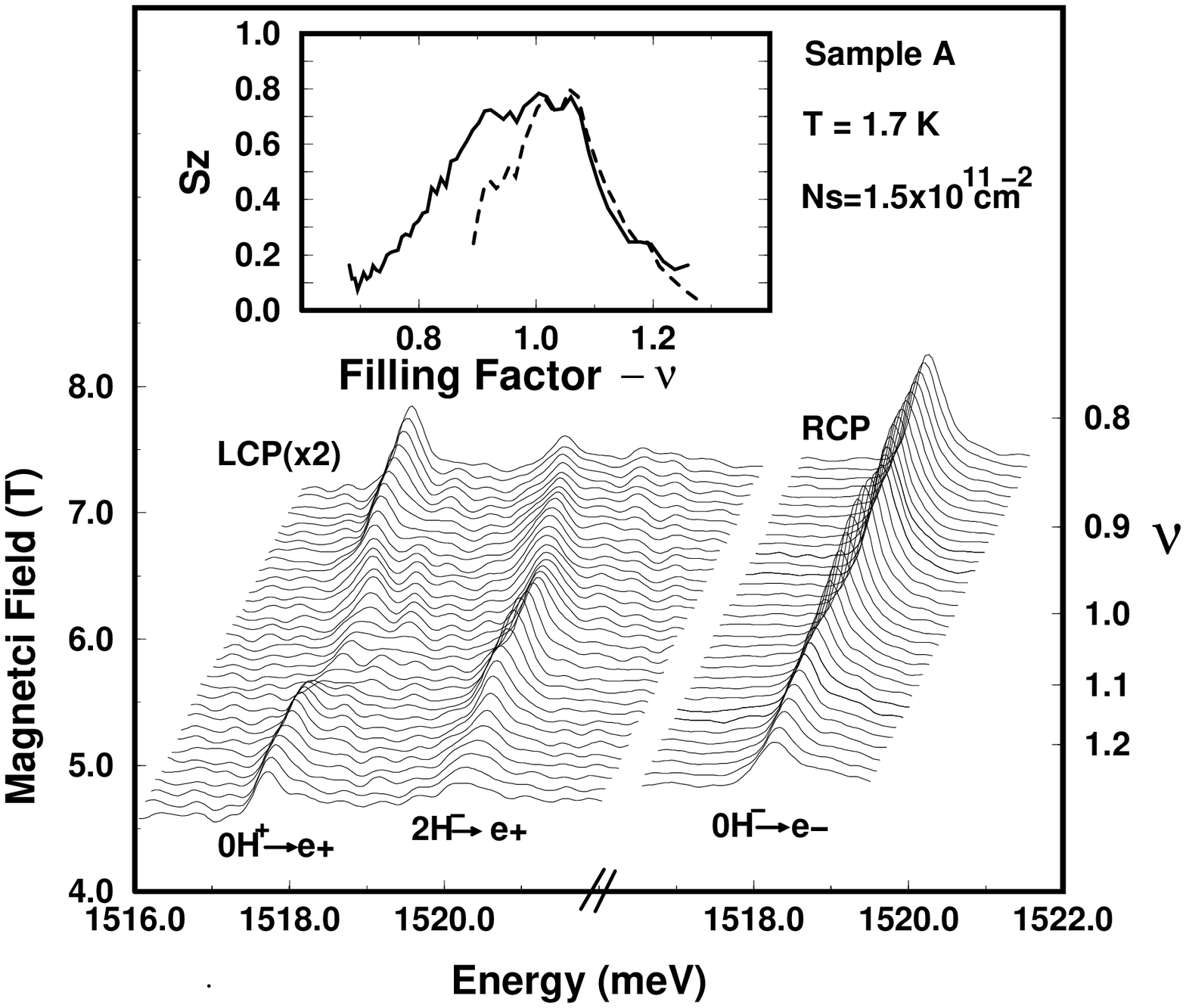,width=6.3in}}
  \begin{figure*}[htbp]
    \begin{center}
      \leavevmode
\centerline{\box\figa}
     \bigskip
      \caption{}
      \label{sky-fig1}
    \end{center}
  \end{figure*}
\vfil
\noindent E. H. Aifer, et al. Figure 1.
\eject
%
%
\newbox\figa
\setbox\figa=\vtop{\kern0pt\psfig{figure=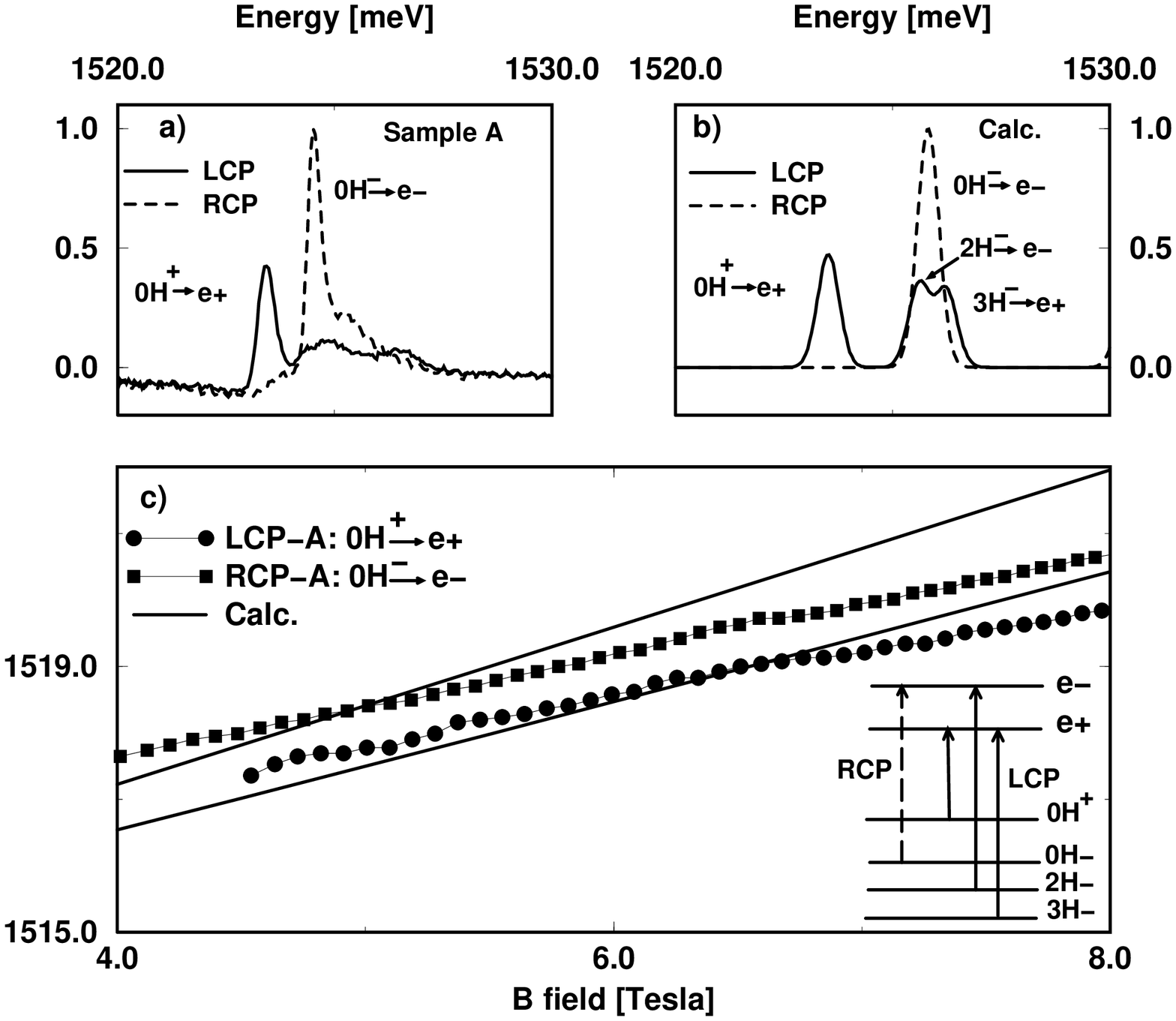,width=6.3in}}
  \begin{figure*}[htbp]
    \begin{center}
      \leavevmode
\centerline{\box\figa}
      \bigskip
      \caption{}
      \label{sky-fig2}
    \end{center}
  \end{figure*}
\vfil
\noindent E. H. Aifer, et al. Figure 2.
\eject
%

%
%
\newbox\figa
\setbox\figa=\vtop{\kern0pt\psfig{figure=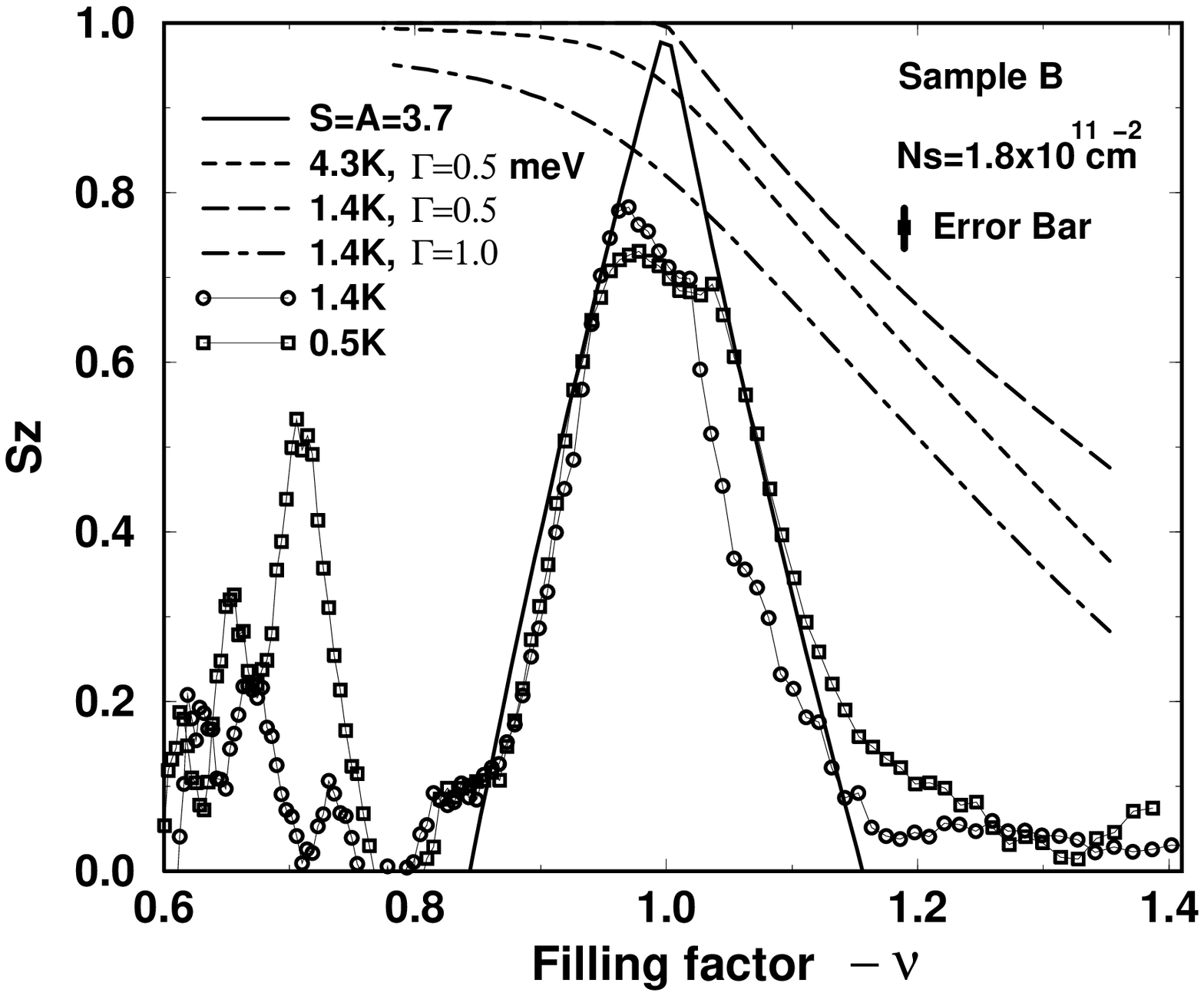,width=6.3in}}
  \begin{figure*}[htbp]
    \begin{center}
      \leavevmode
\centerline{\box\figa}
      \bigskip
      \caption{}
      \label{sky-fig3}
    \end{center}
  \end{figure*}
\vfill\noindent E. H. Aifer, et al. Figure 3.
\eject
%

%
\newbox\figa
\setbox\figa=\vtop{\kern0pt\psfig{figure=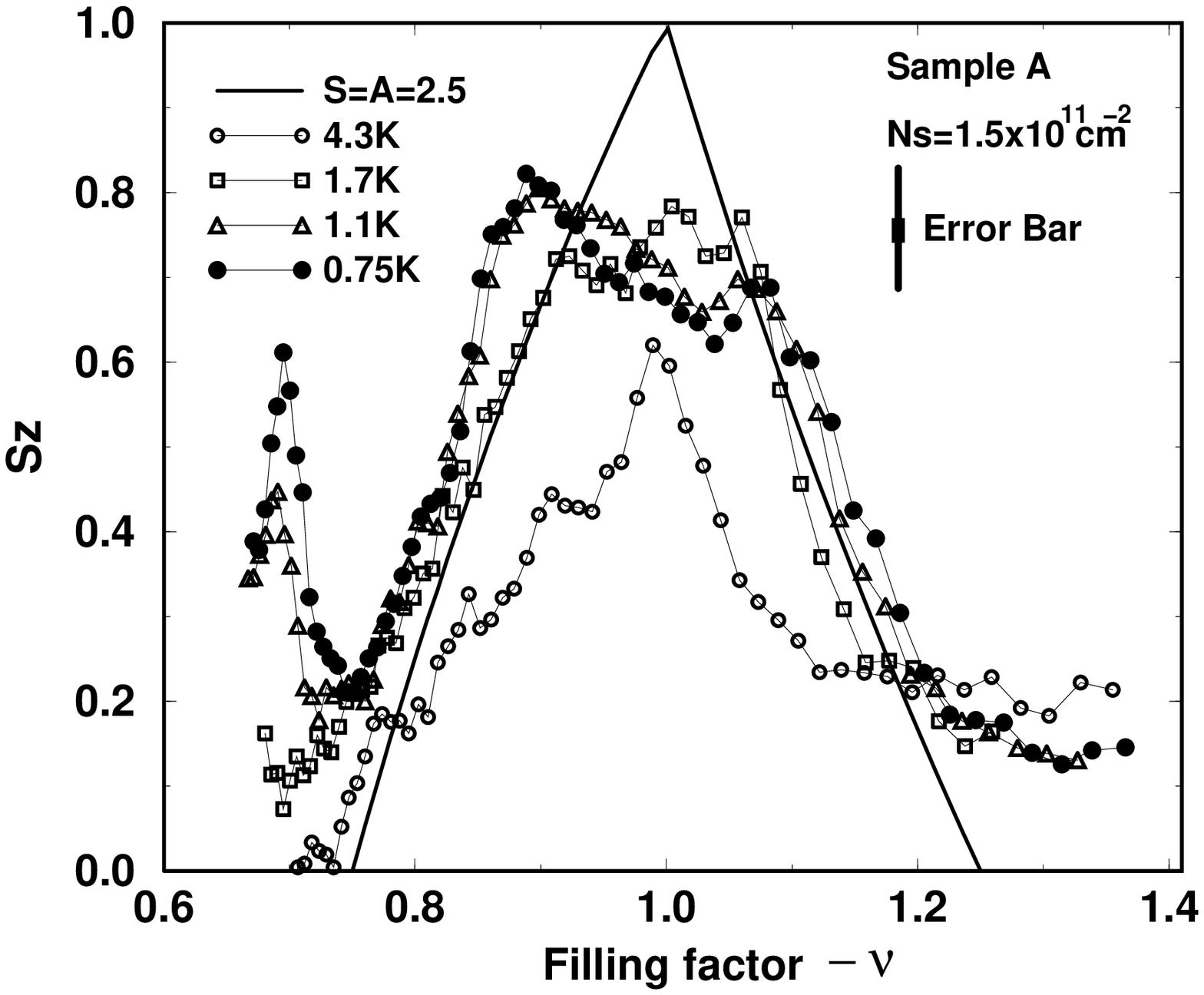,width=6.3in}}
  \begin{figure*}[htbp]
    \begin{center}
      \leavevmode
\centerline{\box\figa}
      \bigskip
      \caption{}
      \label{sky-fig4}
    \end{center}
  \end{figure*}
\vfill
\noindent E. H. Aifer, et al. Figure 4.
\eject
\end{document}